%% file: local-congruence.tex
\documentclass[]{article}
\usepackage{graphicx}				
\usepackage{geometry}                		
\usepackage{amsmath}
\usepackage{hyperref}
\usepackage{txfonts}
\usepackage{lipsum}
\input{macros}

\usepackage{caption}

\usepackage{inconsolata} 
\usepackage[usenames,dvipsnames]{color} 
\usepackage{listings}

\lstdefinelanguage{julia}
{
  keywordsprefix=\@,
  morekeywords={
    exit,whos,edit,load,is,isa,isequal,typeof,tuple,ntuple,uid,hash,finalizer,convert,promote,
    subtype,typemin,typemax,realmin,realmax,sizeof,eps,promote_type,method_exists,applicable,
    invoke,dlopen,dlsym,system,error,throw,assert,new,Inf,Nan,pi,im,begin,while,for,in,return,
    break,continue,macro,quote,let,if,elseif,else,try,catch,end,bitstype,ccall,do,using,module,
    import,export,importall,baremodule,immutable,local,global,const,Bool,Int,Int8,Int16,Int32,
    Int64,Uint,Uint8,Uint16,Uint32,Uint64,Float32,Float64,Complex64,Complex128,Any,Nothing,None,
    function,type,typealias,abstract
  },
  sensitive=true,
  morecomment=[l]{\#},
  morestring=[b]',
  morestring=[b]" 
}

\title{Local congruence of chain complexes}
\author{Gianmaria DelMonte$^a$, Elia Onofri$^b$, Giorgio Scorzelli$^c$ and Alberto Paoluzzi$^b$ \\[3mm]
\small Department of \{Engineering$^a$, Mathematics and Physics$^b$\}, Roma Tre University, Rome, Italy\\
\small Scientific Computing and Imaging Institute$^c$, University of Utah, Salt Lake City, USA}
\date{\today}

\begin{document}
\maketitle

\vspace{-3mm}
\begin{figure}[thbp] 
   \centering
   \includegraphics[width=0.35\linewidth]{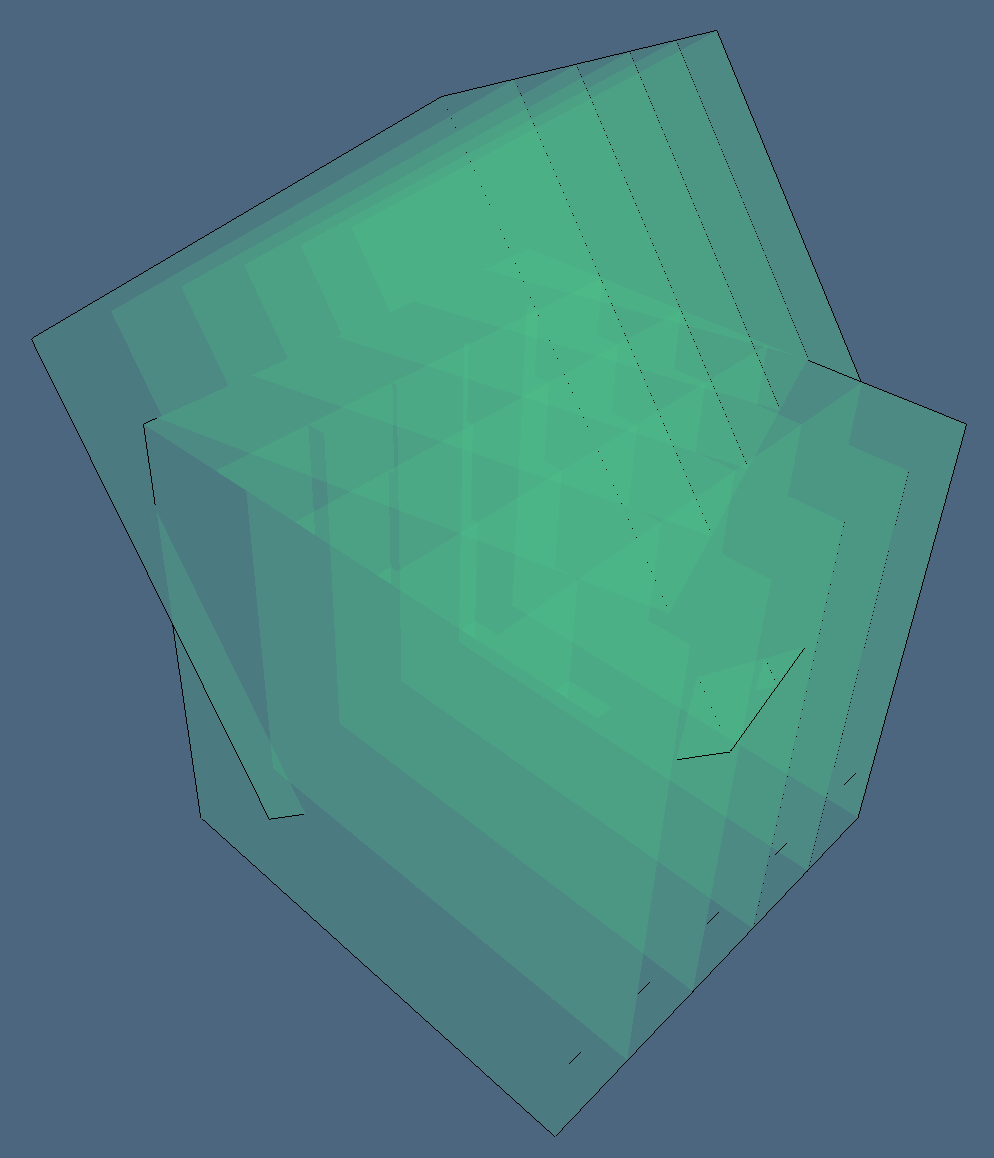}%
   \includegraphics[width=0.65\linewidth]{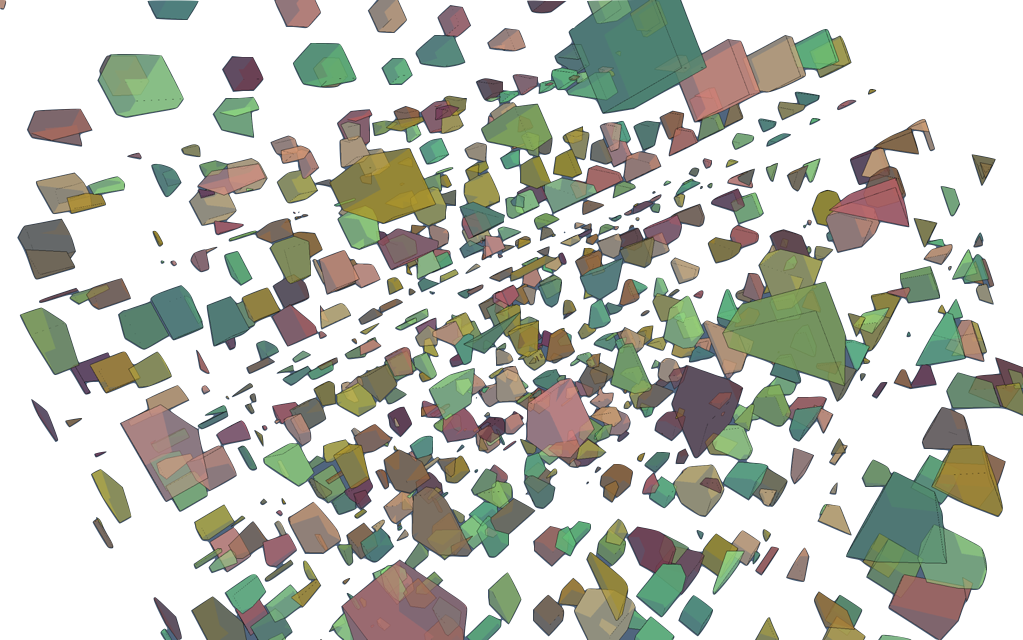}%
   \caption{Cellular 3-complexes: (a) two rotated cuboid grids $5\times 5\times 5$; (b) exploded set of intersected 3-cells. It is interesting to note that  \texttt{\#V} - \texttt{\#E} + \texttt{\#F} = 1192-3182+2824 = 834 = \texttt{\#C}, according to the Euler characteristic $\chi_0 - \chi_1 + \chi_2 - \chi_3 = 0$ of a polyhedral partition of the 3-sphere. The images are generated by the Julia 3D graphics package \href{https://github.com/cvdlab/ViewerGL.jl}{\texttt{https://github.com/cvdlab/ViewerGL.jl}}.}
   \label{fig:example}
\end{figure}

\begin{abstract}
The object of this paper is to transform a set of \emph{local} chain complexes to a single \emph{global} complex using an equivalence relation of $\epsilon$-congruence of cells, solving topologically the numerical inaccuracies of floating-point arithmetics.
While computing the space arrangement generated by a collection of cellular complexes, one may start from independently and efficiently computing the intersection of each single input 2-cell with the others.
The topology of these intersections is codified within a set of (0--2)-dimensional chain complexes. The target of this paper  is to merge the local chains by using the equivalence relations of $\epsilon$-congruence between 0-, 1-,  and 2-cells (elementary chains).
In particular, we reduce the block-diagonal coboundary matrices $[\Delta_0]$ and $[\Delta_1]$, used as matrix accumulators of the \emph{local} coboundary chains, to the \emph{global} matrices $[\delta_0]$ and $[\delta_1]$, representative of congruence topology, i.e., of congruence quotients between all 0-,1-,2-cells, via elementary algebraic operations on their columns. This algorithm is codified using the Julia porting of the \href{http://faculty.cse.tamu.edu/davis/suitesparse.html}{\texttt{SuiteSparse:GraphBLAS}} implementation of the \texttt{GraphBLAS} standard, conceived to efficiently compute algorithms on large graphs using linear algebra and sparse matrices~\cite{Buluc:7965104,GraphBLAS:standard}.
\end{abstract}

\tableofcontents

\section{Introduction}\label{sec:introduction}

Our goal in this paper was to develop and compare different implementations of a computational topology  algorithm needed to solve efficiently and robustly the space decomposition (3D arrangement generated by polyhedral complexes) requested to evaluate Boolean formulas of Constructive Solid Geometry~\cite{2019arXiv191011848P}.

Within the \emph{space decomposition} pipeline to compute the \emph{arrangement}~\cite{Halperin:2017} generated by a collection of cellular complexes~\cite{2017arXiv170400142P,2019arXiv191108130P}, the intermediate step is the assessment of the  $\partial_2$ operator in 3D,
after independent fragmentation of 2-faces in 2D, and before the topological gift wrapping (TGW) algorithm in 3D. 

In particular, the (co)boundary chain is produced by merging a set of local coboundary matrices computed independently from each other on every fragmented input 2-cell. In paper~\cite{2019arXiv191108130P} this computational pipeline is discussed in full detail. In the present manuscript we provide the reduction of a chain complex to its ``quotient topology'' through congruence quotient computation.

We recall that the 2-skeleton of 3D arrangements, and its $[\partial_2]$ matrix, is computed by merging the local chain complexes produced by mutual intersections of 2-cells of input complexes. Each \emph{local} chain complex $C^k_\bullet=(C_p^k, \partial_p^k)$, with $0\leq p\leq 2$, is generated by a  two-dimensional input cell $\sigma^k\in C^k_2$ in the input collection, and by all 2-cells intersecting it. We call \emph{accumulator} complex the set $\varmathbb{C}_\bullet = \{ C_\bullet^k\}$, $1\leq k\leq N$.
Two figures are geometrically \emph{congruent} iff one can be transformed into the other by an isometry~\cite{2017arXiv170400142P}
. The congruences $R_p$ between $p$-cells of geometric complexes in $\varmathbb{C}_\bullet$ are equivalence relations, so to compute the chain complex of  quotient spaces by merging chains and operators by computing the graded quotient $\varmathbb{C}_\bullet/R$.

The merging is based on the discovering of all equivalence classes of the congruence relations between\break 0-, 1-, and 2-cells of decomposed input objects, computed independently~\cite{2017arXiv170400142P} for each input 2-face. 
In~order to discover \emph{all} congruent pairs among decomposed $p$-cells, topological tests are introduced through the computation of local topological
invariants. Our \emph{Cell Congruence Enabling} algorithm may be used to merge the topologies of several cellular complexes (even a large number), after having computed the intersections of their 2-cells. Two examples of actual application of the CCC algorithm are (a) the merging of plan drawings of city districts or buildings within a digital cartographic map, and (b) the mutual fragmentation of cell complexes before computing Boolean expressions~\cite{2019arXiv191011848P} between solid models.

The description of topology, geometry and physics of models using the chain of coboundary operators, i.e., of vector algebra operators, and their representation as sparse matrices, started with~\cite{PALMER1995733,Palmer1993}, at least in the knowledge of the authors. 
While still unusual in the geometric and solid modeling areas (see~\cite{tuprints11291} for example), they seem currently to us the most promising approach to FEA methods~\cite{Desbrun:2006:DDF:1185657.1185665,Elcott:2006:BYO:1185657.1185666,arnold_falk_winther_2006,Arnold:2010,Arnold:2018,Tonti:1975,Tonti:2013,Ferretti:2014}. Also, the GraphBLAS standard, with linear algebra methods based on sparse matrices for fast algorithms on huge graphs, is currently having a big momentum~\cite{graphblas:20}.

The computation of (co)boundary matrices from multiplication of sparse characteristic matrices of $p$-complexes, was introduced in~\cite{Dicarlo:2014:TNL:2543138.2543294}. It may be useful to understand that the chain of (co)boundaries is an algebraic representation of the Hasse diagram of a cellular complex~\cite{10.1145/1236246.1236259,MILICCHIO2008172}. The main advantage of the matrix approach is that topology, geometry and physics may coexist in a unique sparse matrix framework, concurring together to define, represent and simulate the behavior of a model, without any restrictions on type, dimension, codimension, orientability, manifoldness, or connectedness~\cite{10.1145/1236246.1236259}. 

For the sake of reader convenience, the theoretical minimum about linear spaces of chains and their linear operators is given in Section~\ref{sec:chains}. Section~\ref{sec:quotient-topology} introduces the concepts of $\epsilon$-nearness of points and $\epsilon$-congruence of cells in order to define the \emph{quotient topology}.
In Section \ref{sec:coboundaries} we introduce the block-diagonal sparse matrices which set the scene for the Cell Congruence Enabling algorithm, whose Julia code is discussed in~Section \ref{sec:congruence}.
In Section \ref{sec:implementation} both an elegant implementation using native Julia syntax for sparse matrix and vector algebra is given, and its translation to fast GraphBLAS primitives~\cite{graphblas:20} is provided. The concluding section supplies some hints about possible uses of this approach.

\section{Chain spaces and chain complex}\label{sec:chains}

All computer applications including geometric content may require  some computation of topology, i.e., of the relations of incidence or adjacency between \emph{cells} (vertices, edges, faces, volumes), i.e, between 0-, 1-,\break  2-, and 3-cells of a cellular decomposition of either the input objects or their boundaries. 
Cells may be represented as basis elements in a graded vector space of \emph{chains}, closed with respect to addition of cells with the same dimension and w.r.t.~the product times a scalar element in a field~\cite{ieee-tase,Arnold:2018}.

Given a finite collection S of geometric objects in $\E^d$, the arrangement $\mathcal{A}(\mathcal{S})$ is the decomposition of $\E^d$ into connected open cells of dimensions $0, 1, \ldots , d$ induced by $\mathcal{S}$.
We remark that the topology of a space partition is fully described by a \emph{chain complex}, i.e. by a sequence of chain spaces $C_p$ of different dimension ($0\leq p\leq d$), and by linear operators of \emph{boundary} $\partial_p: C_p\to C_{p-1}$ and \emph{coboundary} $\delta_p: C_p\to C_{p+1}$  between chain spaces. In particular, the union of bases of chain spaces provide a minimal set of generators of the induced topology.
Once chosen an ordering of the cells, providing independent chain generators, such linear operators are described by their matrices, holding the scalar values which supply the coordinates with respect to the bases, that are normally either in  $\{0,1\}$ (non-oriented chains) or in $\{-1,0,1\}$ (oriented chains).  

An important problem in computational geometry is to asses the \emph{arrangement}~\cite{Halperin:2017,2019arXiv191108130P} of Euclidean 3-space, i.e., the partition of $\E^3$ produced by a collection of cellular 3-complexes. The simplest computational topology solution is provided by  the computation of the \emph{chain complex} 
\[ 
C_\bullet = (C_p, \partial_p) := 
C_3 \ 
\substack{
\delta_2 \\
\longleftarrow \\[-1mm]
\longrightarrow \\
\partial_3 
}
\ C_2 \ 
\substack{
\delta_1 \\
\longleftarrow \\[-1mm]
\longrightarrow \\
\partial_2 
}
\ C_1 \ 
\substack{
\delta_0 \\
\longleftarrow \\[-1mm]
\longrightarrow \\
\partial_1 
}
\ C_0 .
\] 
of the space partition generated by the input collection of geometric objects. 

The chain of linear boundary operators $(\partial_3, \partial_2, \partial_1)$ is sufficient to characterize the linear chain spaces, since they contain the domain bases by columns, or better, they contain the coordinate vectors representing them as linear combinations of the target space bases. Furthermore, we have $\delta_{p-1} = \partial_p^\top$ by duality, for every $p$.  The computation of the $[\partial_3]$ matrix, via the TGW (Topological Gift Wrapping) algorithm~\cite{2017arXiv170400142P}, requires $[\partial_{2}]$ after the geometrical intersection of input 2-cells. In the present manuscript, we discuss how to compute the matrices $[\delta_0] = [\partial_1]^\top$, and $[\delta_1] = [\partial_2]^\top$, via elementary algebraic operations, implemented by using the standard GraphBLAS library for very large graphs.

\section{Quotient topology}\label{definitions}\label{sec:quotient-topology}

We introduce in this section a simple way to glue together a set of local 
geometric and topological information
generated independently from each input 2-cell, by giving few definitions of 
nearness and congruence, which will allow to transform the union of \emph{local} topologies 
into a single \emph{global} ``quotient topology''.

Let $\epsilon$ be a small positive value. Given a set $V$ of points, we can partition it into subsets $S_i \subset V$ such that $S_i = \{v \in V \mid d(v, v_i) \leq \epsilon\}$ where $v_i$ is a representative of $S_i$, being $d$ the Euclidean distance.
Such partition of a point set, characterized by a clustering of subsets of close elements, can be constructed using the \texttt{inrange} function of the \texttt{NearestNeighbors.jl} package,
which implements the \(k\)-d tree data
structure~\cite{10.1145/361002.361007}. See the Section~\ref{sec:congruence} for details.

\subsection{Nearness}\label{epsilon-nearness}

We say that two points \(u,v\in V\) are \emph{$\epsilon$-near}, and write
\(u\overset{\epsilon}{\sim} v\), when their Euclidean distance is
\(d(u,v) \leq 2\epsilon\).

The \(\epsilon\)-nearness \(\overset{\epsilon}{\sim}\) is an equivalence relation, since it
is reflexive, symmetric, and transitive. In particular, it is transitive since
\emph{any} pair of points \(u,v \in S_i\)
are \(\epsilon\)-near, because both have a distance less than
\(\epsilon\) from \(v_i\), and hence have a distance no more than $2\epsilon$ from each other.
More formally, if \(u\overset{\epsilon}{\sim}v\) and \(v\overset{\epsilon}{\sim}w\), then \(u\overset{\epsilon}{\sim}w\), since the distance  from $v_i$  of every point ({e.g.}, $u,v,w$) in $S_i$ is less than $\epsilon$. 

Let \(W = V/\!\!\overset{\epsilon}{\sim}\) be the quotient set of \(V\)
points w.r.t.~the \(\epsilon\)-nearness relation
\(\overset{\epsilon}{\sim}\), for some fixed \(\epsilon\). The elements
of \(W\) can be chosen as the points \(v_i\) representatives of the equivalence
classes \(S_i\) of \(V\). 
Even better, the representative of each class can be chosen as its $w$ centroid, and the average distance from $w$ of the other points of the class certainly decreases.

\subsection{Topological congruence}\label{epsilon-congruence}

Two geometric figures are \emph{congruent} iff one can be transformed into the other by an isometry~\cite{Coxeter:1967}, i.e.~by an affine transformation with unit determinant.

We say that two \(p\)-cells $e,f$ are \emph{\(\epsilon\)-congruent}, and write $e\cong f$, when there exists a bijection $\mu$ 
between their 0-faces that pairwise maps vertices to \(\epsilon\)-near
vertices. 
The \emph{identification} or \emph{quotient topology} gives a method of getting a
topology on \(X/\!\!\cong\) from a topology on \(X\). The quotient
topology is exactly the one that makes the resulting space `look like'
the original one, with the identified elements glued together.

A subset $B$ of open sets in the topological space \(T\) over the 
cell complex \(X\) is a basis for the topology of \(X\) when all other open 
sets can be written as union of elements of \(B\).
If we have an equivalence relation \(\cong\) on a cell complex \(X\) we get a natural
projection map \(\pi: X \to X/\!\!\cong\), by mapping each cell to its
equivalence class. The \emph{identification} or \emph{quotient topology}  on \(Y = X/\!\!\cong\) is
defined as follows: a set \(A \subset Y\) is open if and only if \(\pi^{-1}(A)\)
is open in \(X\).
In particular, the topology $T(X)$ and the quotient topology $T(Y)$ are equivalent 
because  any non-empty open set of $T(X)$  contains a non-empty open set of $T(Y)$ 
and, conversely, every non-empty open set of $T(Y)$ contains non-empty open sets of $T(X)$.

It is easy to see that \(\epsilon\)-congruence between
elementary chains (aka cells) \(c, d\in C_p\), denoted \(c\cong d\), is a
graded equivalence relation, so that a chain complex \((C_p,\delta_p)\)
may be represented by a much smaller
\(\pi(C_p,\delta_p)= (G_p, \delta'_p)\), where  \(G_p =  \pi(C_p) =  C_p/\!\!\cong\),
and where
\[\delta'_p = \delta_p\circ\pi := \pi(C_p)\to \pi(C_{p+1}) = \delta_p: G_p \to G_{p+1}.\]

\section{Chain Complex Congruence 
}\label{sec:congruence}

In this section we discuss the simple operations needed to transform the initial sparse block matrices $[\Delta_0]$ and $[\Delta_1]$ into the final operator matrices $[\delta_0]$ and $[\delta_1]$ concerning the space partition as a whole.  

As we have seen, a chain $d$-complex is completely specified by the sequence of its boundary or coboundary maps. Hence we intend to construct the coboundary maps induced by the congruence topology on the union set of local topological spaces generated by independent decompositions of input 2-cells, in the computational pipeline~\cite{2017arXiv170400142P} for the building of the space arrangement produced by a collection of cellular complexes.

\begin{figure}[htbp] 
   \centering
   \includegraphics[width=0.35\textwidth]{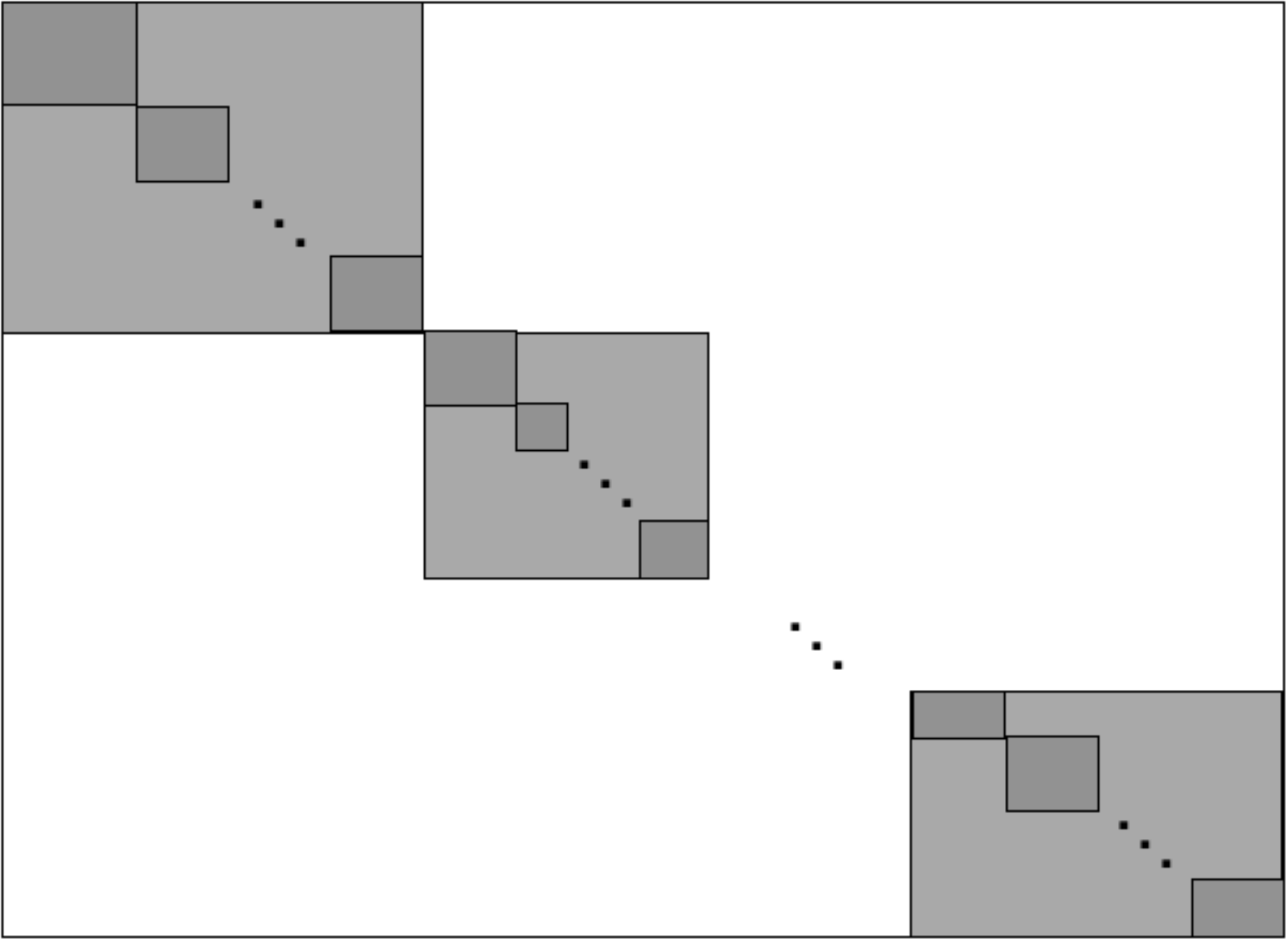} 
   \caption{Doubly nested structure of sparse block-matrices $[\Delta_0]:C_0\to C_1$ and
   $[\Delta_1]:C_1\to C_2$. }
   \label{fig:blocks}
\end{figure}

\subsection{Block-diagonal accumulators of coboundaries}\label{sec:coboundaries}

We fix our attention here to $d=3$ and in particular to the computation of $[\delta_1]$ and $[\delta_0]$. The matrix $[\delta_2] = [\partial_3]^\top$ will be computed by \emph{Topological Gift Wrapping} (TGW) algorithm, not described in this paper, in a following stage of the computational pipeline~\cite{2017arXiv170400142P}, starting from $[\partial_2] = [\delta_1]^t$.

In the algorithmic specification below, 
a dense array  \texttt{W} and two sparse arrays \texttt{Delta\_0}, \texttt{Delta\_1} respectively provide the input vertex coordinates and the sparse block-matrices $[\Delta_0]$ and $[\Delta_1]$.

After independent computation, possibly in parallel, of local chain complexes (see~\cite{2017arXiv170400142P}), both \emph{accumulator matrices} $[\Delta_1]:C_1\to C_2$ and $[\Delta_0]:C_0\to C_1$ have a sparse block-diagonal structure, made by two nested levels of (sparse) diagonal blocks. Each \emph{outer block} concerns one of the $m$ input geometric objects, and $n_k$ \emph{inner blocks}, $1\leq k\leq m$, store the matrices of each decomposed $\xi(\sigma)$ 2-cell. 
 We call $[\theta_h], 1\leq h \leq m$, the exterior blocks (light gray), and $[\epsilon_k], 1\leq k \leq m_h$, the interior blocks (dark gray), where $m_h$ is the number of 2-cells in $k$-th input geometric object.

\subsection{Nearness of vertices}\label{sec:aa:1}

First we need to discover the \emph{$\epsilon$-nearness} of vertices on the $3\times n$ input matrix \texttt{V} of 3D coordinates, by querying the \texttt{kdtree} data structure~\cite{10.1145/361002.361007}, and getting $m$ classes of $\epsilon$-congruent vertices, represented by the array of arrays \texttt{vclasses}. The $3\times m$ output matrix \texttt{W} holds by columns the coordinates of identified vertices in each $\epsilon$-congruent class, that are mapped to their centroids.

\begin{lstlisting} 
function vcongruence(V::Matrix; epsilon=1e-6)
	vclasses, visited = [], []
	kdtree  = NearestNeighbors.KDTree(V);
	for vidx = 1 : size(V, 2)  if !(vidx in visited)
		nearvs = NearestNeighbors.inrange(kdtree, V[:,vidx], epsilon)
		push!(vclasses, nearvs)
		append!(visited, nearvs)  end
	end
	W = hcat([sum(V[:,class], dims=2)/length(class) for class in vclasses]...)
	return W, vclasses
end
\end{lstlisting}

\subsection{\emph{Chain Complex Congruence} Algorithm}\label{sec:aa:2}

In Section~\ref{sec:coboundaries} we have discussed the block diagonal marshaling $[\Delta_0]$ and $[\Delta_1]$ of local coboundary matrices. With the function \texttt{cellcongruence} we replace each subset of columns of \texttt{Delta} sparse matrix  corresponding to \emph{$\epsilon$-near vertices} with their vector sum. In this way we produce a new matrix from an array of new vectors. Finally, equal rows of this new matrix, discovered by a dictionary, are substituted by a single representative. 

The same function \texttt{cellcongruence} is also applied to $[\Delta_1]$, by summing each subset of columns corresponding to each class of \emph{congruent edges}, so generating a new sparse matrix from the resulting set of columns. Then, we reduce every subset of equal rows, if any, to a single row representative of congruent faces. 
A stepwise description of the \emph{Chain Complex Congruence} (CCC) algorithm is given below.  
The function \texttt{cellcongruence} takes  \texttt{Delta} of type \texttt{SparseMatrixCSC} and  the \texttt{inclasses} given as array of arrays, and  computes the set \texttt{outclasses}  of congruent elementary $d$-chains, represented as arrays of $(d-1)$-chains. 

\begin{enumerate}

\item \texttt{Lar.cop2lar} transforms the sparse matrix $[\Delta_p]$ into a \texttt{cellarray} (array of arrays of cell indices); then a vector \texttt{newcell} is defined to accomodate a map between old facet indices into the new ones of their quotient projection $\pi$. Finally, the array \texttt{cells} accomodates the translated representations (arrays) of input ($p$-1)-cells corresponding to columns of $\texttt{Delta}:C_{p-1}\to C_p$;

\item the empty \texttt{cells}, if any, are removed from \texttt{okcells}. Empty cells appear when they   contain a number of \emph{distinct} facets  $n=\texttt{length(Set(cell))}$ less than \texttt{dim}, i.e., $n< 2$ for edges, and $n< 3$ for faces;

\item each $p$-cell (elementary $p$-chain) is represented as an array of indices of its facets ($p$-1)-chain).
The selection of single instances of output cells (from \texttt{okcells}, where they may be repeated) is performed using a \texttt{DefaultOrderedDict} for congruence \texttt{classes}, with key given by the cell representation (array of facets). The dictionary reading action \texttt{classes[face]} returns \texttt{[]} iff the key (\texttt{face}) is not contained in the dictionary. In this case the key is stored, and its storage of value starts with the first index \texttt{[k]}; otherwise, the current index is appended to the yet incomplete value. 

\item At the end, both the dictionary \emph{keys} (column indices of congruent rows, i.e.~the set of  non-zero positions (for the representative row) in the output $[\delta_p]$, and the dictionary \emph{values}, i.e., the projected elements (class representatives) are given as output.

\end{enumerate}

\begin{lstlisting}[mathescape=true] %xxxxxxxxxxxxxxxxxxxxxxxxxxxxxxxxxx

function cellcongruence(Delta, inclasses; dim)
  cellarray = Lar.cop2lar(Delta)
  newcell = Vector(undef, size(Delta,2))
  [ newcell[e] = k for (k, class) in enumerate(inclasses) for e in class ]
  cells = [map(x -> newcell[x], face) for face in cellarray]
  okcells = [cell for cell in cells if length(Set{cell)) $>$ dim]  # non-empty cells
  classes = DefaultOrderedDict{Vector, Vector}([])
  for (k,face) in enumerate(okcells)
    classes[face] == [] ?  classes[face] = [k] : append!(classes[face], [k])
  end
  cells = collect(keys(classes))
  outclasses = collect(values(classes))
  return cells, outclasses
end
\end{lstlisting} 

\subsection{Top-level interface}\label{sec:aa:3}

Finally, the higher-level function \texttt{chaincongruence} maps the input data \texttt{W}, \texttt{Delta\_0}, \texttt{Delta\_1}, into a compact representation \texttt{V}, \texttt{EV}, \texttt{FE} of the chain complex $\texttt{V}:C_0\to\E^3$, $\delta_0:C_0\to C_1$, and $\delta_1:C_1\to C_2$ (see Appendix~\ref{sec:appendix-1}).
For full generality, also the 2-cell congruence was checked and taken into account. Congruent faces would in fact appear when some input 2-cells of the arrangement pipeline have inner intersection and lay on the same 3D plane. 
We would like to remark that decomposed 1-cells (i.e., input edges) and the representatives of their congruence classes (i.e., output edges) are associated one-to-one with the rows of $[\Delta_0]\to[\delta_0]$ and the columns of $[\Delta_1]\to[\delta_1]$, respectively, so satisfying the topological constraints $[\Delta_1][\Delta_0]=[\delta_1][\delta_0]=[0]$, that were checked positively in all our tests.

\begin{lstlisting} %xxxxxxxxxxxxxxxxxxxxxxxxxxxxxxxxxx

function chaincongruence(W, Delta0, Delta_1)
  V, vclasses = vcongruence(W)
  EV, eclasses = cellcongruence(Delta_0, vclasses, dim=1)
  FE, fclasses = cellcongruence(Delta_1, eclasses, dim=2)
  return V, EV, FE
end

V,EV,FE = chaincongruence(W,Delta_0,Delta_1)
\end{lstlisting} 
\label{page:chaincongruence}

\section{Sparse matrix implementations}\label{sec:implementation}

Two compact matrix implementations of Chain Complex Congruence algorithm are given in this section, starting from native coding in Julia, with its concise matrix syntax, then using the Julia porting of the package \texttt{SuiteSparse:GraphBLAS}, enriched by a simple user-friendly interface from Julia's multiple dispatch.
The Julia code for the three implementations of the CCC algorithm is available at open-source repository \href{https://github.com/cvdlab/LocalCongruence.jl}{\texttt{https://github.com/cvdlab/LocalCongruence.jl}}.
The three implementations are denoted as \texttt{AA} (array of arrays), \texttt{SP} (Julia sparse array), and \texttt{GB} (GraphBLAS), respectively. The \texttt{AA} code was discussed in Section~\ref{sec:congruence}. 

\subsection{Julia's \texttt{SparseArrays.jl} implementation}\label{ceaSM}

The \texttt{vertexCongruence} evaluates the \emph{Vertex Congruence} for 3D-points. The function determines the points of $3\times n$ matrix ``V`` closer than ``err`` to each other, and builds a new \emph{Vertex Set} made of the representative of each point cluster.
The method returns: the new Vertex Set; a map that, for every new vertex, identifies the subset of old vertices it is made of.
The function cellCongruence evaluates the \emph{Cell Congruence} for a $d$-Cochain ``cop`` of type sparse array, with $d-1$ equivalence classes ``lo\_cls``, and corresponding ``lo\_sign``, for $(d-1)$-cells, both given as array of arrays of (old) lower-rank indices, to denote equivalence classes. The \texttt{chainCongruence} function performs the Geometry ``G`` congruence and reshapes the Topology ``T``.

\subsection{Mapping to \texttt{SuiteSparse:GraphBLAS.jl}}\label{ceaGB}

\section{Analysis of results}\label{sec:conclusion}

We have introduced here three versions of our CCC algorithm. A comparison with similar or different methods is clearly not possible, by the novelty of our approach, but a mutual comparison is interesting.
The first algorithm, in Section~\ref{sec:congruence}, introduced the sparse block-matrix reduction method by using arrays of arrays, but with the drawback of losing the ordering information (and the sparse matrix output) of the cells, which were available in the input. The second version is given  in Section~\ref{ceaSM} by directly using the Julia's native sparse matrices. The third implementation, in Section~\ref{ceaGB}, makes use of the GraphBLAS primitives within a tiny Julia's wrapping. 

Below you may see, using mean times, and assuming \texttt{AA = 1x} (Native Julia's array of array --- last column), that we have:
native Julia (first column) \texttt{SparseArrays = 16.6x;} Julia wrapping (second column) of \texttt{SuiteSparse:GraphBLAS = 6.12x}.

\hspace{-5mm}
{\scriptsize
\begin{minipage}{0.33\textwidth}
\begin{lstlisting}[mathescape=true]
julia> @benchmark chainCongruence(W,T)
BenchmarkTools.Trial: 
  memory estimate:  19.47 MiB
  allocs estimate:  782635
  --------------
  minimum time:     1.006 s (0.38% GC)
  median time:      1.008 s (0.00% GC)
  mean time:        1.008 s (0.16% GC)
  maximum time:     1.010 s (0.00% GC)
  --------------
  samples:          5
  evals/sample:     1
\end{lstlisting}
\end{minipage}
\begin{minipage}{0.36\textwidth}
\begin{lstlisting}[mathescape=true]
julia> @benchmark chainCongruenceGB(W,T_GB)
BenchmarkTools.Trial: 
  memory estimate:  30.07 MiB
  allocs estimate:  1077403
  --------------
  minimum time:     363.749 ms (0.00% GC)
  median time:      371.598 ms (1.06% GC)
  mean time:        370.236 ms (0.80% GC)
  maximum time:     378.384 ms (1.20% GC)
  --------------
  samples:          14
  evals/sample:     1
\end{lstlisting}
\end{minipage}
\begin{minipage}{0.35\textwidth}
\begin{lstlisting}[mathescape=true]
julia> @benchmark chainCongruenceAA(G,T)
BenchmarkTools.Trial: 
  memory estimate:  9.33 MiB
  allocs estimate:  154347
  --------------
  minimum time:     55.680 ms (0.00% GC)
  median time:      59.759 ms (0.00% GC)
  mean time:        60.497 ms (1.66% GC)
  maximum time:     71.103 ms (7.29% GC)
  --------------
  samples:          83
  evals/sample:     1
\end{lstlisting}
\end{minipage}
}

All implementations are quite naive. No optimizations were performed. We believe that major benefits may be ported by optimization to the two versions that use sparse matrices. A great \emph{caveat} for the array of arrays version is that it looses track of cell signs, that must be recovered later with further computations, so losing all time benefits. Anyway, it let us suppose that the GraphBLAS approach can be forther optimized.

The two implementations with sparse matrices allow for maintaining the 2-cell orientation within the global chain complex. We have seen that the \texttt{GraphBLAS} implementation has two important benefits:
(1)~with parity of storage occupation, it allows to compute the largest chain complexes; (2) it also allows for block decomposed matrix computations, so opening the way to hierarchical computation of huge topological operators. Finally, we remark that the CCC algorithm is multidimensional, and its implementation may be extended, \emph{as is}, to manage higher dimensional complexes, just by adding more \texttt{cellcongruence} call instances to \texttt{chaincongruence} function on page~\pageref{page:chaincongruence}.

\section{Conclusion}\label{sec:conclusion}

In this paper we have discussed an efficient algebraic algorithm to create the $[\delta_0]$ and $[\delta_1]$ sparse coboundary matrices, encoding the topological congruences between a set of chain complexes. In the algorithmic pipeline to construct the arrangement of Euclidean 3-space~\cite{2019arXiv191108130P}, local chain complexes are generated independently from single fragmented input 2-cells, starting from a collection of cellular 2- or 3-complexes in 3D. 
The correctness of results, that might depend on numeric approximations of floating-point arithmetics, is  checked by testing the matrix constraint $[\delta_1][\delta_0]=[0]$.
Therefore, we have given one possible topological solution to the robustness problem in geometric computations, in the line discussed by Christoph Hoffmann in~\cite{Hoffmann:2001}, using epsilon geometry~\cite{10.1145/73833.73857}.
The chain complex congruence (CCC) enabling algorithm introduced here was inplemented in Julia using the package \href{https://github.com/abhinavmehndiratta/SuiteSparseGraphBLAS.jl}{\texttt{SuiteSparseGraphBLAS.jl}}, and it is a key component of a computational pipeline to produce solid models of complex geometric scenes, using robust Boolean algebra methods for next-generation image understanding.

\bibliographystyle{IEEEtran}
\bibliography{ccomplexblas.bib}

\appendix
\section{Appendix}
\subsection{A simple example} \label{sec:appendix-1}

For the sake of reproducibility, as well as of reader convenience and understanding, a very simple example of the congruence reduction for a 3D cube is given here, by using the initial implementation of Sections~\ref{sec:aa:1} and~\ref{sec:aa:2}. The open-source realization with Julia sparse matrices and/or with GraphBLAS may by downloaded from GitHub at  \texttt{
 \href{https://github.com/cvdlab/LocalCongruence.jl}{https://github.com/cvdlab/LocalCongruence.jl}}.
 
The cube was generated with random size and random attitude produced by a random 3D rotation, in order to start from close but unequal coordinates for each single instance of face vertices. All faces were generated independently, in order to get $6\times 4 = 24$ columns (vertex instances) in the input matrix \texttt{W}. The Julia's matrix input follows (as $24\times 3$ matrix, in order to show the full \texttt{Float64} digital resolution).

\begin{lstlisting} %xxxxxxxxxxxxxxxxxxxxxxxxxxxxxxxxxx

julia> W = convert(Matrix,[0.5310492999999998 0.8659989999999999 0.14191280000000003; 1.0146684 0.6827212999999999 0.2169682; 0.3477716 0.5268921 0.4947971000000001; 0.8313907882395298 0.3436144447063971 0.5698524407571428; 0.6061046999999998 1.2188832999999994 0.5200012; 1.0897237999999998 1.0356056999999999 0.5950565999999999; 0.42282699999999984 0.8797763999999998 0.8728855; 0.9064461979373597 0.6964987903021808 0.9479408896095312; 0.5310493 0.8659989999999999 0.14191279999999987; 1.0146684000000001 0.6827213 0.21696819999999994; 0.6061047 1.2188833 0.5200011999999999; 1.0897237772434623 1.035605657895053 0.5950566438156151; 0.3477716 0.5268921 0.4947971; 0.8313908 0.3436145 0.5698525000000001; 0.422827 0.8797764000000001 0.8728855; 0.9064462 0.6964988 0.9479409; 0.5310493 0.8659989999999999 0.14191280000000006; 0.34777160000000007 0.5268920999999999 0.4947971000000001; 0.6061047 1.2188833 0.5200012000000002; 0.4228270000000002 0.8797764 0.8728855000000001; 1.0146684 0.6827213 0.21696819999999994; 0.8313908 0.3436145 0.5698525000000001; 1.0897238 1.0356057 0.5950565999999999; 0.9064461675456482 0.6964988122992563 0.9479408949632379]'); 
\end{lstlisting} 
The two corresponding sparse block-diagonal matrices \texttt{Delta\_0} and \texttt{Delta\_1} are given below, as array triples \texttt{(I,J,X)} of row and column indices, and non-zero values. 

\begin{lstlisting} %xxxxxxxxxxxxxxxxxxxxxxxxxxxxxxxxxx

julia> Delta_0 = SparseArrays.sparse([1, 3, 1, 4, 2, 3, 2, 4, 5, 7, 5, 8, 6, 7, 6, 8, 9, 11, 9, 12, 10, 11, 10, 12, 13, 15, 13, 16, 14, 15, 14, 16, 17, 19, 17, 20, 18, 19, 18, 20, 21, 23, 21, 24, 22, 23, 22, 24], [1, 1, 2, 2, 3, 3, 4, 4, 5, 5, 6, 6, 7, 7, 8, 8, 9, 9, 10, 10, 11, 11, 12, 12, 13, 13, 14, 14, 15, 15, 16, 16, 17, 17, 18, 18, 19, 19, 20, 20, 21, 21, 22, 22, 23, 23, 24, 24], Int8[-1, -1, 1, -1, -1, 1, 1, 1, -1, -1, 1, -1, -1, 1, 1, 1, -1, -1, 1, -1, -1, 1, 1, 1, -1, -1, 1, -1, -1, 1, 1, 1, -1, -1, 1, -1, -1, 1, 1, 1, -1, -1, 1, -1, -1, 1, 1, 1]);
\end{lstlisting} 

\begin{lstlisting} %xxxxxxxxxxxxxxxxxxxxxxxxxxxxxxxxxx

julia> Delta_1 = SparseArrays.sparse([1, 1, 1, 1, 2, 2, 2, 2, 3, 3, 3, 3, 4, 4, 4, 4, 5, 5, 5, 5, 6, 6, 6, 6], [1, 2, 3, 4, 5, 6, 7, 8, 9, 10, 11, 12, 13, 14, 15, 16, 17, 18, 19, 20, 21, 22, 23, 24], Int8[1, -1, -1, 1, 1, -1, -1, 1, 1, -1, -1, 1, 1, -1, -1, 1, 1, -1, -1, 1, 1, -1, -1, 1]);
\end{lstlisting} 
The execution command from Julia terminal is given below, followed by output showing, with \texttt{V} output by columns, and both \texttt{EV} (edges by vertex indices) and \texttt{FE} (faces by edge indices) as array of arrays:

\begin{lstlisting} %xxxxxxxxxxxxxxxxxxxxxxxxxxxxxxxxxx

julia> V,EV,FE = chaincongruence(W,Delta_0,Delta_1);
\end{lstlisting} 

\begin{lstlisting}[mathescape=true] %xxxxxxxxxxxxxxxxxxxxxxxxxxxxxxxxxx

julia> @show V;  # centroids of the 8 $\overset{\epsilon}{\sim}$ classes of W points
V = [0.531049 1.01467  0.347772 0.831391 0.606105 1.08972  0.422827 0.906446; 
     0.865999 0.682721 0.526892 0.343614 1.21888  1.03561  0.879776 0.696499; 
     0.141913 0.216968 0.494797 0.569852 0.520001 0.595057 0.872886 0.947941]
\end{lstlisting} 

\hspace{-6.5mm}
\begin{minipage}[l]{0.475\linewidth}
\begin{lstlisting} %xxxxxxxxxxxxxxxxxxxxxxxxxxxxxxxxxx

julia> @show EV;  # edges-by-vertices
EV = Array{T,1} where T[[1, 2], [1, 3], [1, 5], [2, 4], [2, 6], [3, 4], [3, 7], [4, 8], [5, 6], [5, 7], [6, 8], [7, 8]]
\end{lstlisting} 

\begin{lstlisting} %xxxxxxxxxxxxxxxxxxxxxxxxxxxxxxxxxx

julia> @show FE;  # faces-by-edges
FE = Array{T,1} where T[[1, 2, 3, 4], [1, 5, 9, 10], [2, 6, 11, 12], [3, 7, 9, 11], [4, 8, 10, 12], [5, 6, 7, 8]]
\end{lstlisting} 
\end{minipage}
\hfill
\begin{minipage}[]{0.525\linewidth}
   \centering
   \includegraphics[width=0.45\linewidth]{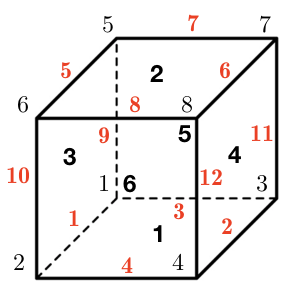} 
   \captionof{figure}{The generated chain complex, with basis 0-chains (roman), 1-chains (red) and 2-chains (bold).}
   \label{fig:example}
\end{minipage}

\end{document}

%% file: macros.tex
\def\E{\mathbb{E}}


%% file: local-congruence.bbl
\begin{thebibliography}{10}
\providecommand{\url}[1]{#1}
\csname url@samestyle\endcsname
\providecommand{\newblock}{\relax}
\providecommand{\bibinfo}[2]{#2}
\providecommand{\BIBentrySTDinterwordspacing}{\spaceskip=0pt\relax}
\providecommand{\BIBentryALTinterwordstretchfactor}{4}
\providecommand{\BIBentryALTinterwordspacing}{\spaceskip=\fontdimen2\font plus
\BIBentryALTinterwordstretchfactor\fontdimen3\font minus
  \fontdimen4\font\relax}
\providecommand{\BIBforeignlanguage}[2]{{%
\expandafter\ifx\csname l@#1\endcsname\relax
\typeout{** WARNING: IEEEtran.bst: No hyphenation pattern has been}%
\typeout{** loaded for the language `#1'. Using the pattern for}%
\typeout{** the default language instead.}%
\else
\language=\csname l@#1\endcsname
\fi
#2}}
\providecommand{\BIBdecl}{\relax}
\BIBdecl

\bibitem{Buluc:7965104}
\BIBentryALTinterwordspacing
A.~Buluc, T.~Mattson, S.~McMillan, J.~Moreira, and C.~Yang, ``{Design of the
  GraphBLAS API for C},'' in \emph{2017 IEEE International Parallel and
  Distributed Processing Symposium: Workshops (IPDPSW)}.\hskip 1em plus 0.5em
  minus 0.4em\relax Los Alamitos, CA, USA: IEEE Computer Society, jun, 2017,
  pp. 643--652. [Online]. Available:
  \url{https://doi.ieeecomputersociety.org/10.1109/IPDPSW.2017.117}
\BIBentrySTDinterwordspacing

\bibitem{GraphBLAS:standard}
------, ``{The GraphBLAS C API Specification},'' Tech. Rep., 2019.

\bibitem{2019arXiv191011848P}
\BIBentryALTinterwordspacing
A.~{Paoluzzi}, V.~{Shapiro}, A.~{DiCarlo}, G.~{Scorzelli}, and E.~{Onofri},
  ``{Finite Boolean Algebras for Solid Geometry using Julia's Sparse Arrays},''
  \emph{arXiv e-prints}, p. arXiv:1910.11848, Oct. 2019. [Online]. Available:
  \url{https://ui.adsabs.harvard.edu/abs/2019arXiv191011848P}
\BIBentrySTDinterwordspacing

\bibitem{Halperin:2017}
D.~Halperin and M.~Sharir, ``Arrangements,'' in \emph{Handbook of Discrete and
  Computational Geometry -- Third Edition}, J.~E. Goodman, J.~O'Rourke, and
  C.~D. T\`oth, Eds.\hskip 1em plus 0.5em minus 0.4em\relax Boca Raton, FL,
  USA: CRC Press, Inc., 2017, ch.~28.

\bibitem{2017arXiv170400142P}
\BIBentryALTinterwordspacing
A.~{Paoluzzi}, V.~{Shapiro}, and A.~{DiCarlo}, ``{Regularized arrangements of
  cellular complexes},'' \emph{arXiv e-prints}, p. arXiv:1704.00142, Apr. 2017.
  [Online]. Available:
  \url{https://ui.adsabs.harvard.edu/abs/2017arXiv170400142P}
\BIBentrySTDinterwordspacing

\bibitem{2019arXiv191108130P}
\BIBentryALTinterwordspacing
A.~{Paoluzzi}, V.~{Shapiro}, A.~{DiCarlo}, F.~{Furiani}, G.~{Martella}, and
  G.~{Scorzelli}, ``{Topological computing of arrangements with (co)chains},''
  \emph{arXiv e-prints}, p. arXiv:1911.08130, Nov. 2019. [Online]. Available:
  \url{https://ui.adsabs.harvard.edu/abs/2019arXiv191108130P}
\BIBentrySTDinterwordspacing

\bibitem{PALMER1995733}
\BIBentryALTinterwordspacing
R.~S. Palmer, ``Chain models and finite element analysis: An executable chains
  formulation of plane stress,'' \emph{Computer Aided Geometric Design},
  vol.~12, no.~7, pp. 733 -- 770, 1995, {Grid Generation, Finite Elements, and
  Geometric Design}. [Online]. Available:
  \url{http://www.sciencedirect.com/science/article/pii/016783969500015X}
\BIBentrySTDinterwordspacing

\bibitem{Palmer1993}
\BIBentryALTinterwordspacing
R.~S. Palmer and V.~Shapiro, ``Chain models of physical behavior for
  engineering analysis and design,'' \emph{Research in Engineering Design},
  vol.~5, no.~3, pp. 161--184, Sep 1993. [Online]. Available:
  \url{https://doi.org/10.1007/BF01608361}
\BIBentrySTDinterwordspacing

\bibitem{tuprints11291}
\BIBentryALTinterwordspacing
J.~S. Mueller-Roemer, ``Gpu data structures and code generation for modeling,
  simulation, and visualization,'' Ph.D. dissertation, Technische
  Universit{\"a}t Darmstadt, Darmstadt, 2020. [Online]. Available:
  \url{http://tuprints.ulb.tu-darmstadt.de/11291/}
\BIBentrySTDinterwordspacing

\bibitem{Desbrun:2006:DDF:1185657.1185665}
\BIBentryALTinterwordspacing
M.~Desbrun, E.~Kanso, and Y.~Tong, ``Discrete differential forms for
  computational modeling,'' in \emph{ACM SIGGRAPH 2006 Courses}, ser. SIGGRAPH
  '06.\hskip 1em plus 0.5em minus 0.4em\relax New York, NY, {USA}: Acm, 2006,
  pp. 39--54. [Online]. Available:
  \url{http://doi.acm.org/10.1145/1185657.1185665}
\BIBentrySTDinterwordspacing

\bibitem{Elcott:2006:BYO:1185657.1185666}
\BIBentryALTinterwordspacing
S.~Elcott and P.~Schroder, ``Building your own dec at home,'' in \emph{ACM
  SIGGRAPH 2006 Courses}, ser. SIGGRAPH '06.\hskip 1em plus 0.5em minus
  0.4em\relax New York, NY, {USA}: Acm, 2006, pp. 55--59. [Online]. Available:
  \url{http://doi.acm.org/10.1145/1185657.1185666}
\BIBentrySTDinterwordspacing

\bibitem{arnold_falk_winther_2006}
D.~N. Arnold, R.~S. Falk, and R.~Winther, ``Finite element exterior calculus,
  homological techniques, and applications,'' \emph{Acta Numerica}, vol.~15, p.
  1–155, 2006.

\bibitem{Arnold:2010}
------, ``Finite element exterior calculus: from {H}odge theory to numerical
  stability,'' \emph{Bull. Amer. Math. Soc. (N.S.)}, vol.~47, pp. 281--354,
  2010.

\bibitem{Arnold:2018}
D.~N. Arnold, \emph{Finite Element Exterior Calculus}, ser. CBMS-NSF Regional
  Conference Series in Applied Mathematics.\hskip 1em plus 0.5em minus
  0.4em\relax Philadelphia, PA: Society for Industrial and Applied Mathematics
  (SIAM), 2018, vol.~93.

\bibitem{Tonti:1975}
E.~Tonti, ``On the formal structure of physical theories,'' Italian National
  Research Council, Tech. Rep., 1975.

\bibitem{Tonti:2013}
------, \emph{The Mathematical Structure of Classical and Relativistic
  Physics}.\hskip 1em plus 0.5em minus 0.4em\relax Birkh\"auser, 2013.

\bibitem{Ferretti:2014}
E.~Ferretti, \emph{The Cell Method: A Purely Algebraic Computational Method in
  Physics and Engineering}.\hskip 1em plus 0.5em minus 0.4em\relax Momentum
  Press, 2014.

\bibitem{graphblas:20}
\BIBentryALTinterwordspacing
``Graph blas forum,'' 2014-2020, reference implementations. [Online].
  Available: \url{http://graphblas.org/index.php?title=Graph_BLAS_Forum}
\BIBentrySTDinterwordspacing

\bibitem{Dicarlo:2014:TNL:2543138.2543294}
\BIBentryALTinterwordspacing
A.~DiCarlo, A.~Paoluzzi, and V.~Shapiro, ``Linear algebraic representation for
  topological structures,'' \emph{Comput. Aided Des.}, vol.~46, pp. 269--274,
  Jan. 2014. [Online]. Available:
  \url{http://dx.doi.org/10.1016/j.cad.2013.08.044}
\BIBentrySTDinterwordspacing

\bibitem{10.1145/1236246.1236259}
\BIBentryALTinterwordspacing
A.~DiCarlo, F.~Milicchio, A.~Paoluzzi, and V.~Shapiro, ``Solid and physical
  modeling with chain complexes,'' in \emph{Proceedings of the 2007 ACM
  Symposium on Solid and Physical Modeling}, ser. SPM ’07.\hskip 1em plus
  0.5em minus 0.4em\relax New York, NY, {USA}: Association for Computing
  Machinery, 2007, p. 73–84. [Online]. Available:
  \url{https://doi.org/10.1145/1236246.1236259}
\BIBentrySTDinterwordspacing

\bibitem{MILICCHIO2008172}
\BIBentryALTinterwordspacing
F.~Milicchio, A.~DiCarlo, A.~Paoluzzi, and V.~Shapiro, ``A codimension-zero
  approach to discretizing and solving field problems,'' \emph{Advanced
  Engineering Informatics}, vol.~22, no.~2, pp. 172 -- 185, 2008, {Network}
  methods in engineering. [Online]. Available:
  \url{http://www.sciencedirect.com/science/article/pii/S1474034607000468}
\BIBentrySTDinterwordspacing

\bibitem{ieee-tase}
\BIBentryALTinterwordspacing
A.~DiCarlo, F.~Milicchio, A.~Paoluzzi, and V.~Shapiro, ``Chain-based
  representations for solid and physical modeling,'' \emph{Automation Science
  and Engineering, {IEEE} Transactions on}, vol.~6, no.~3, pp. 454 --467, july
  2009. [Online]. Available: \url{https://doi.org/10.1109/TASE.2009.2021342}
\BIBentrySTDinterwordspacing

\bibitem{10.1145/361002.361007}
\BIBentryALTinterwordspacing
J.~L. Bentley, ``Multidimensional binary search trees used for associative
  searching,'' \emph{Commun. {ACM}}, vol.~18, no.~9, p. 509–517, Sep. 1975.
  [Online]. Available: \url{https://doi.org/10.1145/361002.361007}
\BIBentrySTDinterwordspacing

\bibitem{Coxeter:1967}
H.~S.~M. Coxeter and S.~L. Greitzer, \emph{Geometry Revisited}.\hskip 1em plus
  0.5em minus 0.4em\relax Washington, D.C.: Math. Assoc. Amer., 1967.

\bibitem{Hoffmann:2001}
C.~M. Hoffmann, ``Robustness in geometric computations,'' Computer Science,
  Purdue University, Tech. Rep., April 1, 2001.

\bibitem{10.1145/73833.73857}
\BIBentryALTinterwordspacing
D.~Salesin, J.~Stolfi, and L.~Guibas, ``Epsilon geometry: Building robust
  algorithms from imprecise computations,'' in \emph{Proceedings of the Fifth
  Annual Symposium on Computational Geometry}, ser. SCG'89.\hskip 1em plus
  0.5em minus 0.4em\relax New York, NY, {USA}: Association for Computing
  Machinery, 1989, pp. 208â--217. [Online]. Available:
  \url{https://doi.org/10.1145/73833.73857}
\BIBentrySTDinterwordspacing

\end{thebibliography}
